\documentclass[a4paper]{article}

\usepackage[pages=all, color=black, position={current page.south}, placement=bottom, scale=1, opacity=1, vshift=5mm]{background}
\SetBgContents{
	
}      

\usepackage[margin=1in]{geometry} 
\usepackage{color,soul}
\usepackage{amsmath}
\usepackage{amsthm}
\usepackage{amssymb}
\usepackage{hyperref}
\hypersetup{
	unicode,
	pdfauthor={Shayan Roofeh, Vahid Karimipour},
	pdftitle={Capacities of Generalized Landau-Streater Channel},
	pdfsubject={Capacities of Generalized Landau-Streater Channel},
	pdfkeywords={article, template, simple},
	pdfproducer={LaTeX},
	pdfcreator={pdflatex}
}

\usepackage[sort&compress,numbers,square]{natbib}
\usepackage{bm}
\theoremstyle{plain}

\theoremstyle{definition}

\usepackage{graphicx}
\graphicspath{{fig/}}
\usepackage{subcaption}
\usepackage{algorithm, algpseudocode} 
\usepackage{mathrsfs} 
\usepackage{lipsum}
\usepackage{physics}
\def\be{\begin{equation}}
	\def\ee{\end{equation}}
\def\ba{\begin{eqnarray}}
	\def\ea{\end{eqnarray}}
\def\lo{\longrightarrow}
\def\h{\hskip 1cm }

\def\la{\langle}
\def\ra{\rangle}
\def\ni{\noindent}
\def\a{\alpha}

\def\bex{\begin{dinglist}{110}\dsquare}
	\def\eee{\end{dinglist}}
\def\bet{\begin{dinglist}{110}\bsquare}
	\def\bfr{\begin{mdframed}[backgroundcolor=blue!20]\vspace{0.5cm}}
		\def\efr{\vspace{0.5cm}\end{mdframed}}
	
	\title{ The noisy Landau-Streater (Werner-Holevo) channel in arbitrary dimensions}
	\author{ Vahid Karimipour}
	
	\date{
		Department of Physics, Sharif University of Technology, Tehran, Iran\\%
	}
	
\begin{document}
		\maketitle
		
		\begin{abstract}
			Two important classes of quantum channels, namly the Werner-Holevo and the Landau-Streater channels are known to be related only in three dimensions, i.e. when acting on qutrits.  In this work, definition of the Landau-Streater channel  is extended in such a  way which retains its equivalence to the Werner-Holevo channel in all dimensions. This channel is then modified to be representable as a model of noise acting on qudits.   We then investigate  propeties of the resulting noisy channel and determine the conditions under which it cannot be the result of a Markovian evolution. Furthermore, we investigate its different capacities for transmitting classical and quantum information with or without entanglement.  In particular, while the pure (or high noise) Landau-Streater or the Werner-Holevo channel is entanglement breaking and hence has zero capacity, by finding a lower bound for the quantum capacity, we show that when the level of noise is lower than a critical value the quantum capacity will be non-zero. Surprizingly this value turns out to be approximately  equal to $0.4$ in all dimensions.  Finally we show that, in even dimension,  this channel has a decomposition in terms of unitary operations. This is in contrast with the three dimensional case where it has been proved that such a decomposition  is impossible, even in terms of other quantum maps. \\

		\end{abstract}

	\section{Introduction} 
	High-dimensional quantum channels have a wide range of practical applications across different areas of quantum information science, offering advantages such as increased information capacity \cite{Datta_2005, smith2010quantum, shor_2003,Chessa_2021}, enhanced security \cite{Islam_2017}, \cite{Karimipour_2002}, and improved performance in quantum technologies \cite{Wang_2020}. There has also been advances in concrete realization of high dimensional quantum systems in various physical systems. Notable among these are the encoding of a $d-$ level system or qudit in the angular momentum of structured light \cite{Rubinsztein-Dunlop_2017}, \cite{Gao_2024},\cite{Karimi:12},\cite{PhysRevApplied.11.064058}. Apart from these practical considerations, there has always been strong interest in extending the formalism of quantum information beyond two level systems or the so-called qubits. It is important to explore the limitations and powers of high-dimensional quantum states and quantum channels. Notable among these are the generalized Pauli channels \cite{Siudzinska_2020},\cite{Akio_Fujiwara_2003} and multi-level amplitude damping channels\cite{Chessa_2021}. Interesting examples of high dimensional quantum channels, include the Werner Holevo \cite{WH} channels and the Landau-Streater channels \cite{LanS}.  The former is an example of an entanglement-breaking channel,  which destroys all the entanglement in the input state, the study of which provides insight into the capacities of other quantum channels in general. The latter, while also being entanglement breaking  is an example of an extreme point in the space of all quantum channels, i.e. a channel which cannot be represented as the convex combination of other channels. Moreover it is an example of a unital channel which cannot be represented as a mixture of unitary operations, contrary to the qubit case, where this is always possible \cite{Audenaert_2008}. \\
	
	For these and other reasons, particularly  their symmetries, these two types of channels have attracted a lot of attention in quantum information science. 
		 Various properties of the Landau-Streater channel have been studied for example in \cite{Audenaert_2008, filippov_LS, pakhomchik_realization_2020} and those of the Werner-Holevo channel in \cite{ Girard_2022, datta2004additivity, Cope_2017, Cope_2018, Chitambar_2023, Wolf_2005, fannes2004additivity}.	 Below we will remind the reader of their definition.\\
	
\noindent	{\bf The  Landau Streater (LS) channel:}  Let the dimension be $d=2j+1$, where $j$ is an integer or half-integer, then the Landau-Streater (LS) channel for qudits (acting on density matrices belonging to a d-dimensional Hilbert space) is defined as 
	
		\be\label{WHLS}
	{\Lambda}_j(\rho)=\frac{1}{j(j+1)}(J_x\rho J_x+J_y\rho J_y+J_z\rho J_z)
	\ee
	where $J_x, J_y$ and $J_z$ are the spin-$j$ representation of generators of rotations in 3-dimensional space, commonly called the $so(3)$ algebra. These generators are Hermitian and satisfy the algebra $[J_a,J_b]=i\epsilon_{a,b,c}J_c$ . In a  Hilbert space $V={\rm Span}\{|j,m\ra, m=-j\cdots j\}$, they are represented as 
		\ba
		J_z|j,m\ra&=&m|j,m\ra\cr
		J_{\pm}|j,m\ra&=& \sqrt{j(j+1)-m(m\pm 1)}|j,m\pm1\ra
		\ea
		where $J_\pm=J_x\pm i J_y.$  This is the first example  \cite{LanS}, of a unital quantum channel which cannot be realized as a collection of random unitary operations. More concretely, the map ${\cal L}_j$, while having the property ${\cal L}_j(I)=I$, cannot be written as ${\cal L}_j(\rho)=\sum_i p_i U_i \rho U_i^\dagger$ for any choice of unitary actions and any choice of randomness. Moreover, this channel  is an extreme point in the space of quantum channels. This is an intriguing result since it is well known that for qubits, any unital map can be written as a random unitary channel \cite{konrad}.  This means that the LS channel cannot model an environmental noise in any way and should be looked at solely as a mathematical and abstract model. In other words, there is no parameter which can be tuned to represent the level of noise by which we can interpolate between the identity channel and the LS channel. For obvious  reasons and for emphasizing what will be defined in the sequel,  we call this the $SO(3)$ Landau-Streater channel or $SO(3)$ LS channel for short. \\

\noindent		{\bf The Werner-Holevo (WH) channel:} On the same Hilbert space as above, a Werner-Holevo (WH) channel \cite{WH}  is defined as \cite{Cope_2017}
			\be\label{whcope}
			\phi(\rho):= \frac{1}{d-1}\left[\tr\rho\  \mathcal{I}-\rho^T)\right]
			\ee
			\ni (Actually we are here dealing with a specific channel among the one-parameter family of Werner-Holevo channels.)
 This is an example of a quantum channel with  entanglement-breaking property \cite{Horodecki_2003} and was used as a counterexample of the additivity of minimal output R\'enyi entropy \cite{datta2004additivity, WH, fannes2004additivity}. This channel has the covariance property under $SU(d)$ group, that is:
 \be
 \phi (U\rho U^\dagger)=U^*\phi (\rho)U^T,\h \forall\ U\in U(d),
 \ee
 a property which facilitates many calculations relating to the capacities of quantum channels. \\
 
 \ni {\bf Recent works:} In view of the importance of the two channels, a natural question arises whether or not they are related in any way. The answer is known to be positive for the so-called qutrits, i.e. for $3$-level systems, when $d=3$. Therefore when  $j=1$, it is known that the two channels are the same, \cite{filippov_LS, pakhomchik_realization_2020,roofeh2023noisy}, that is: 
\be
\frac{1}{2}(J_x\rho J_x+J_y\rho J_y+J_z\rho J_z)= \frac{1}{2}\left[\tr\rho\  \mathcal{I}-\rho^T)\right]
\ee
To show this equivalence, one of course needs to use a specific representation of the spin-1 representation of $so(3)$, namely,
  \begin{align}\label{KrausLS}
	J_x &= -i \begin{bmatrix} 0 & 0 & 0 \\ 0 & 0 & 1 \\ 0 & -1 & 0 \end{bmatrix}, &
	J_y &= -i \begin{bmatrix} 0 & 0 & -1 \\ 0 & 0 & 0 \\ 1 & 0 & 0 \end{bmatrix}, &
	J_z &= -i \begin{bmatrix} 0 & 1 & 0 \\ -1 & 0 & 0 \\ 0 & 0 & 0 \end{bmatrix},
\end{align}
		otherwise the equivalence is established up to unitary conjugation \cite{filippov_LS}. The equivalance of the two channels however, stops at this point, namely in dimension three (i.e. for qutrits).\\
		
		 \ni Note that the LS (or equivalently the Werner-Holevo channel) cannot be used as models of noisy channels where the parameter of noise can be tuned, i.e. to implement a low-noise channel. In \cite{roofeh2023noisy}, it was shown that one can modify this channel as follows
\be
\Lambda_x(\rho):=(1-x)\rho + x \Lambda_1(\rho)\h 0\leq x \leq 1,
\ee
	and this new channel can indeed act as a noisy Landau-Streater or  Werner-Holevo channel. Moreover it was shown that this channel allows a simple physical interpretation in the form of random rotations, that is 
	\be
	\Lambda_x(\rho)=\int d{\bf n}d\theta P({\bf n},\theta)e^{i{\bf n}\cdot {\bf J}\theta}\rho e^{-i{\bf n}\cdot {\bf J}\theta},
	\ee
	where $x$ is related to the probability distribution $P({\bf n},\theta)$ \cite{roofeh2023noisy}.
  It was then shown in \cite{roofeh2023noisy} how various capacities of the modified channel, being a rather feasible model of quantum noise on qutrits \cite{caves_qutrit_2000, brus_optimal_2002,molina-terriza_experimental_2005,kendon_bounds_2002, cerf_greenberger-horne-zeilinger_2002,bartlett_quantum_2002, bouda_entanglement_2001}, can be calculated or lower- and upper-bounded. Most interestingly, it was analytically shown in \cite{roofeh2023noisy} that the channel $\Lambda_x$ is anti-degradable if the parameter $x$ is greater than a critical value $x_c=\frac{4}{7}$. This means that the  quantum capacity of this channel is exactly zero beyond this critical value. If we regard $x$ as the noise parameter, this means that when the level of noise is higher than this $x_c=\frac{4}{7}$, no quantum information can be sent through this channel in any reliable way, no matter how and by  how much redundancy we encode or decode the quantum states. \\

\noindent Quite recently, this  analysis was taken one step further, when Lo, Lee and Hsieh \cite{lo2024degradability}, studied the $SO(3)$ Landau Streater channel in arbitrary dimensions and its noisy verion  (i.e. for higher-spin representation but of the three-dimensional rotation group). They tackled the problem of degradability and showed that 
in the low noise regime, these channels are $O(\epsilon^2)$ degradable. We remind the reader that a channel $\Phi$ is degradable \cite{devetak_capacity_2005,cubitt_structure_2008} if it is related by another Completely Positive Trace-Preserving (CPT) map to its complement, namely if there is a CPTP map $\Psi$, such that $\Phi^c=\Psi\circ \Phi.$ $\epsilon$-degradability \cite{Sutter_2017}, means that this equality is valid only approximately, that is $\Vert \Phi^c-\Psi\circ \Phi\Vert_{\diamond}=O(\epsilon) $, where $\Vert\cdot \Vert_{\diamond}$ is the diamond norm.   This result narrows down the value of quantum capacity, which usually cannot be calculated exactly. \\

\noindent {\bf The present work:} In this paper we proceed to do the following:\\

\ni {\bf a)}  As mentioned above, the identity of the Landau-Streater and the Werner-Holevo channel stops at the spin-1 representation of the $SO(3)$ group. By considering the  higher spin representations of the rotation group, we are still in the realm of the $SO(3)$ Landau-Streater and there is no equivalence with the Werner-Holevo channel. To make this connection,  we replace the $SO(3)$ group with $SO(d)$, the group of rotations in $d$ dimensional space, which is naturally shown to be equivalent to the $d$-dimensional Werner-Holevo channel.\\

\ni {\bf b)} In the same way as in \cite{roofeh2023noisy}, we make a convex combination of this channel with an identity channel to construct a one-parameter family of channels in the form 
\be\label{deff}
\Phi_{_{x}}(\rho)=(1-x)\rho + \frac{x}{d-1}(\tr(\rho)\ I -\rho^T).
\ee
\noindent Thus in any dimension $d$, we are dealing with a one-parameter family of channels. We then study several properties of this channel, Namely we\\

{\bf b1)} characterize the full spectrum of the channel, which determines the range of the parameter $x$, where this channel cannot be infinitesimally divided and hence cannot be the result of a Markovian evolution.  \\

{\bf b2)} determine  the one-parameter family of complementary channels $\Phi^c_{x}$ in closed form,\\

{\bf b3)} show that in even dimensions, the Werner Holevo or the Landau-Streater channel and its noisy extension has a mixed unitary representation, that is, we prove that in these dimensions, the channel can be written as a convex combination of unitary operations. To the best of our knowledge this is a property not known for the Werner-Holevo channel, \\

\ni {\bf c)} calculate its classical one-shot capacity in  the full parameter range, and the entanglement-assisted capacity in closed form. Furthermore we determine a lower bound for the  quantum capacity and show that  there is a critical value of $x_0$, below which the channel $\Phi_x$ has definitely a non-zero quantum capacity. The value of this critical parameter turns out to be approximately equal to $0.4$ in all dimensions. Above this value of noise, we do not know if the capacity of the channel is zero or not. \\

	\noindent The structure of this paper roughly corresponds to the points a, b, b1,b2,b3 and c discussed above. We end the paper with a discussion. \\
	
	\noindent{{\bf Remark on notation:}} Throughout the paper, we use the notation $\Lambda$ for the standard Landau-Streater channel based on the group $SO(3)$, $\Phi$ for our definition of the $SO(d)$ Landau-Streater channel and ${\cal E}$ for an arbitrary channel. \\

\section{Definition of the $SO(d)$ Landau-Streater channels}\label{def}
Let $R^d$ denote the $d-$ dimensional Cartesian space and let $\mathcal{H}_d$ be a Hilbert space of dimension $d$ with basis states $\{|n\ra, n=1\cdots d\}$. The space of linear operators on $\mathcal{H}_d$ is denoted by $\mathcal{L}({\cal H}_d)$ and the set of positive linear operators on $\mathcal{H}_d$ by $\mathcal{L}^+({\cal H}_d)$. The density operators on this Hilbert space is denoted by $\mathcal{D}({\cal H}_d)$. The operators 
	\be\label{JJ}
J_{mn}=-i(|m\ra\la n|-|n\ra\la m|)\h 
\ee		
are the generators of the Lie algebra $SO(d)$, the Lie algebra of the group $SO(d)$ or rotations in $R^d$.  $J_{mn}$ generates rotations in the $m-n$ plane in $R^d$. The set of operators $\Delta_-:=\{J_{mn}, 1\leq m<n\leq d\}$, is indeed closed under commutation relations    
\be
[J_{kl},J_{mn}]=i\{\delta_{lm} J_{kn}+\delta_{kn}J_{lm}-\delta_{ln} J_{km}+\delta_{km}J_{ln}\},
\ee		
showing that $so(d)$ is indeed a Lie-algebra of dimension $\frac{d(d-1)}{2}$. Furthermore one can also see that 
\be
\sum_{m< n}J_{mn}^\dagger J_{mn}=(d-1)\ \mathcal{I}.
\ee		
By taking $J_{mn}$ to be the Kraus operators of a map, one can then  define a completely positive trace-preserving quantum map or quantum channel which turns out to be 
\be
\Phi(\rho):=\frac{1}{(d-1)}\sum_{m<n}J_{mn}\rho J_{mn}^\dagger = \frac{1}{d-1}\left[\tr\rho\  I-\rho^T)\right]
\ee
To prove this, it is better to use the anti-symmetry of the Kraus operators $J_{mn}=-J_{nm}$ and write

\ba\label{symm}
\Phi(\rho)&:=&\frac{1}{2(d-1)}\sum_{m,n}J_{mn}\rho J_{mn}^\dagger \cr 
&=&\frac{1}{2(d-1)}\sum_{m,n}(|m\ra\la n|-|n\ra\la m|)\rho (|n\ra\la m|-|m\ra\la n|) \cr
&=& \frac{1}{(d-1)}\sum_{m,n}(\rho_{n,n}|m\ra\la m|-\rho_{m,n}|n\ra\la m|) \cr
&=&\frac{1}{d-1}\left[\tr\rho\  \mathcal{I}-\rho^T)\right]
\ea
This is the generalization of the well-known $SO(3)$ Landau-Streater channel to  arbitrary dimensions. We call it the {\bf $SO(d)$ Landau Streater channel}. \\

\noindent {\bf Remark:} {\it Hereafter we use the names Werner-Holevo (WH) channel and Landau-Streater (LS) channels interchangably. }\\

\noindent The last equality shows that it is equivalent to the Werner-Holevo channel,equation (\ref{whcope}). While the Kraus operators belong to the 
algebra of $SO(d)$, the resulting channel is covariant under the full group of unitary matrices $U(d)$, that is
\be
\Phi(U\rho U^\dagger)=U^*\phi^{-}(\rho)U^T\h U\in U(d).
\ee
This channel, while of great interest, is not yet appropriate to model a noisy quantum channel, since there is no term which interpolates this to the identity channel. We can now add such a term and define a one-parameter channel as 
\be
\Phi_{_{x}}(\rho)=(1-x)\rho + \frac{x}{d-1}(\tr(\rho)\ \mathcal{I}-\rho^T).
\ee
 We call this the noisy Landau-Streater or the noisy Werner-Holevo channel. 
 Note however that the addition of the identity channel now leads to a reduction of the covariance group from  $U(d)$, the group of all $d-$ dimensional unitaries to its subgroup $O(d)$, the group of all orthogonal matrices, for which $U=U^*$, 
\be
\Phi_{_{x}}(U\rho U^\dagger)=U\Phi_{_{x}}(\rho)U^\dagger,\h U=U^* \in O(d)
\ee
 We are now prepared to study the spectrum of the channel $\phi_{_x}$. 

\section{Spectrum of the channel and its infinitesimal divisibility $\phi_{_x}$}\label{spectrum}
It is an interesting question as to when a given quantum channel is the result of a Markovian evolution or even when it is infinitesimally divisible. When the Landau-Streater channel is mixed with the identity channel to model an environmental noise, this question becomes relevant for the resulting channel $\Phi_x$. An interesting result of  \cite{wolf_dividing_2008} gives an answer to this question in the negative sense, that is it states that if the determinant of a channel is negative, then the channel is not infinitesimally divisible. Therefore we calculate the determinant of the channel $\Phi_x$.\\

 \ni Consider the matrices $E_{ij}=|i\ra\la j|$ and let 
\ba
&&X_{ij}:=E_{ij}+E_{ji}\h i\leq j \cr
&&Y_{ij}:=E_{ij}-E_{ji}\h i<j \cr
&&Z_{i}:=E_{ii}-E_{i+1,i+1}.
\ea
We first derive the spectrum of the channel $\phi_-$. It is a matter of direct calculation to verify the following relations, where in each case, $g$ denotes the degeneracy of a given eigenvalue.  
\begin{align}
	\Phi_{_{x}}(X_{ij})&=\big[1-x\frac{d}{d-1}\big]X_{ij} &g&=\frac{d(d-1)}{2}\cr
	\Phi_{_{x}}(Y_{ij})&=\big[1-x\frac{d-2}{d-1}\big]Y_{ij} &g&=\frac{d(d-1)}{2}\cr
	\Phi_{_{x}}(Z_{i})&=\big[1-x\frac{d}{d-1}\big]Z_{i} & g&=d-1\cr
	\Phi_{_{x}}(I)&=I & g&=1.
\end{align}

\noindent Denoting these eigenvalues by $\lambda_k$, we find the determinant of the channel 
\be\label{det}
Det(\Phi_{_{x}}):=\prod_k \lambda_k=\big[1-x\frac{d}{d-1}\big]^{\frac{(d+2)(d-1)}{2}}\big[1-x\frac{d-2}{d-1}\big]^{\frac{d(d-1)}{2}}
\ee
The negativity of $Det(\Phi_{_{x}})$ depends on the dimension.  From (\ref{det}), it is seen that depending on the dimension $d$, the channel $\Phi_{_{x}}$ is not divisible if:\\

\be
	1<x\frac{d}{d-1},\ \ \ {\rm and}\ \ \ \frac{(d+2)(d-1)}{2}\ {\ \rm is\  odd}.
\ee
That is if
\be
\frac{d-1}{d}<x,\ \ \ {\rm and}\ \ \ d\in \{3,4,7,8,11,12,\cdots\}=\{3+4i, 4+4i, i=1,2,3\cdots\}.
\ee

\ni When the above condition holds, the channel is not the result of a Markovian evolution.

\section{Complementary Channel}
\label{Comp-channel}
The concept of the complement of a channel which is crucial in determining many of the properties of a quantum channel,  hinges on the well-known Stinespring's dilation theorem  \cite{stinespring}, which states that any quantum channel ${\cal E}: A\lo B$ can be constructed as a  unitary map $U:A\otimes E\lo B\otimes E'$, where $E$ and $E'$ are the environments of $A$ and $B$ respectively. More formally, we have 

\begin{equation}
	{\cal E}(\rho) = \text{tr}_{E'}(U\rho U^{\dagger}),
\end{equation}
where $U$ denotes an isometry mapping from $A$ to ${B} \otimes E'$. In this configuration, the complementary channel ${\cal E}^c: A \longrightarrow E'$ is defined by:
\begin{equation} \label{steincomp}
	{\cal E}^c(\rho) = \text{tr}_{B}(U\rho U^{\dagger}),
\end{equation}
constituting a mapping from the input system to the output environment. It's important to note that the complement of a quantum channel is not unique, but there exists a connection between them through isometries, as detailed in \cite{datta_complementarity_2006}. The Kraus operators of the channel ${\cal E}$ and its complement ${\cal E}^c$ are related as follows
\cite{smaczynski2016selfcomplementary}: 
\begin{equation} \label{krauscomp}
	\begin{split}
		&{\cal E}(\rho)= \sum_\alpha A_{\alpha} \rho A_{\alpha} ^{\dagger} \\
		&{\cal E}^c(\rho)= \sum_i R_i \rho R_i ^{\dagger} \\
		&(R_i)_{\alpha,j}= (A_{\alpha})_{i,j}
	\end{split}     
\end{equation}
The last formula gives a very simple recipe for writing the Kraus operators of the complementary channel easily. Put the first rows of all the Kraus operators in consecutive rows of a matrix and call it $R_1$, put the second rows of all the Kraus operators in consecutive rows of a matrix and call it $R_2$, and so on and so forth. To this end, we rewrite the channel $\Phi_x$ as
\be\label{symmcomplement}
\Phi_x(\rho)=(1-x)\rho +\frac{x}{2(d-1)}\sum_{m,n}J_{mn}\rho J_{mn}^\dagger,
\ee
where the summation is over all indices $m$ and $n$. We then write the Kraus operators of the channel in a specific double-index notation, so we write these Kraus operators as
\be\label{theAs1}
A_{0}=\sqrt{1-x}\ I,\ \ \  A_{m,n}=\sqrt{\frac{x}{2(d-1)}}J_{m,n}\ \ \ \ .
\ee
With the number of Kraus operators that we have used to define the channel $\Phi_{{x}}$, $\Phi_{_{x}}^c(\rho)$ will be a square matrix acting on a space $V$ of  dimension $ d^2+1$. This space is partitioned into 
$\ V=V^0\oplus V$, 
where they are respectively spanned by the following normalized vectors

\be\label{v1}
\{|0\ra\}\cup   \{|m,n\ra,\ m,n=1,\cdots d\}.
\ee
The basis vectors of different subspaces are obviously orthogonal to each other and within each subspace, they are orthonormal. In general we can calculate the matrix elements of $\Phi^c_{_{x}}(\rho)$ as follows.  In  view of (\ref{krauscomp}), we have
\ba
\Big[\Phi_{_{x}}^c(\rho)\Big]_{\a,\gamma}&=&\sum_{i,j} (R_i)_{\a,j}\rho_{jk}(R_i^\dagger)_{k,\gamma}\cr
&=&\sum_{i,j} (A_\a)_{i,j}\rho_{jk}(A_\gamma^\dagger)_{k,i}
=\tr(A_\a \rho A_\gamma^\dagger)\h \a,\ \gamma \in \{0,mn\}
\ea
With these conventions and with the explicit expression that we have for $J_{mn}$,  it is readily calculated that 
\be
\big[\Phi_{_{x}}^c(\rho)\big]_{0,0}=\tr(\rho)({1-x})
\ee

\be
\big[\Phi_{_{x}}^c(\rho)\big]_{0,mn}= i\sqrt{\frac{x(1-x)}{2(d-1)}}(\rho_{mn}-\rho_{nm})
\ee
and 
\be
\big[\Phi_{_{x}}^c(\rho)\big]_{mn,pq}=\frac{x}{2(d-1)}\big(\delta_{m,p}\rho_{n,q}-\delta_{n,p}\rho_{mq}-\delta_{m,q}\rho_{np}+\delta_{n,q}\rho_{mp}\big),
\ee
All this can be neatly arranged in a matrix form as follows, where the blocks which from top to bottom and from left to right are spanned by the basis vectors of $V^0$ and $V$ respectively:
\be\label{compblock}
\Phi_{_{x}}^c(\rho)=\begin{pmatrix} (1-x)\tr(\rho) & i\sqrt{\frac{x(1-x)}{2(d-1)}}\la \rho|(I\otimes I-S)\\ 
	- i\sqrt{\frac{x(1-x)}{2(d-1)}}(I\otimes I-S)| \rho\ra&\frac{x}{2(d-1)}(I-S)(I\otimes \rho + \rho\otimes I)\end{pmatrix},\ee
where $|\rho\ra=\sum_{m,n}\rho_{mn}|m,n\ra$ is the vectorized form of $\rho$ and $S$ is the swap operator on $V$, e.g. $S|m.n\ra=|n,m\ra$. 
One can readily check that $\tr(\Phi_{_{x}}^c(\rho))=1$. On can also see that  the off-block diagonal vanish for any diagonal matrix $\rho$. The reason is that the vectorized form of any diagaonal matrix $\rho=\sum_i \lambda_i|i\ra\la i|$ is given by $|\rho\ra=\sum_i \rho_{ii}|i,i\ra$ which is annihilated by the operator $(I\otimes I-S).$

\section{The question of extremality and mixed unitary representation}\label{mix}
The SO(3) Landau-Streater channel, or equivalently the qutrit Werner-Holevo channel 
$\Lambda_{_{WH}}(\rho)=\frac{1}{2}(\tr(\rho)I-\rho^T )$,
originally introduced in \cite{LanS} is known to be an extreme point in the space of quantum channels. That is, it cannot be written as the convex combination of other CPT maps.  One can see this simply by noting the well-known theorem of Choi \cite{LanS} according to which a CPT map 
$${\cal E}(\rho)=\sum_{m=1}^K A_m\rho A_m^\dagger,$$
is extremal if and only if the set $\{A_m^\dagger A_n,\ m,n=1\cdots K\}$ is linearly independent. For the SO(3) channel, where the Kraus operators belong to the set $\{J_x, J_y, J_z\}$, this is obviously true. The question arises whether this is also the case for the $SO(d)$ channel. As we will show below, it turns out that for higher groups $SO(d)$, this is no longer the case.   Moreover, the $SO(3)$ channel has the property that while it is a unital, it cannot be represented by a mixed unitary channel, i.e. a channel whose Kraus operators are unitary matrices. We will see that contrary to this case, at least for the $SO(2d)$ case, the Landau-Streater or the Werner-Holevo channel can be decomposed in terms of unitary operations.  \\

\ni To prove non-extremality, it is enough to invoke the Choi theorem \cite{LanS} and note that when the non-ordered pair of indices are such that $$\{m,n\}\ne \{p,q\},$$ then $J_{mn}J_{pq}=0$ and there are many such pairs when $d\geq 4$, which makes these pairs linearly dependent. For example in $d=4$, the pairs $J_{12}J_{34},\ 
J_{13}J_{24},$ and $J_{14}J_{23}$ all vanish  and are hence linearly dependent.\\

\ni  A more interesting question is whether or not, such a channel has a mixed unitary representation, the answer to which is positive, at least when $d$ is even. In order  not to clutter the notation, we describe the basic idea by two simple examples, namely $d=4$ and $d=6$. The reasoning easily generalizes to $d=2k$. \\

\subsection{A mixed unitary representation for $SO(4)$ Landau-Streater channel}
Let us construct a different set of Kraus operators for this channel in the form 
\be
K^{\pm}_1=\frac{1}{\sqrt{2}}(J_{12}\pm J_{34}),\ \ \ \ K^{\pm}_2=\frac{1}{\sqrt{2}}(J_{13}\pm J_{24}),\ \ \ \ K^{\pm}_3=\frac{1}{\sqrt{2}}(J_{14}\pm J_{23}).
\ee
It is easily seen that  ${K^\pm_i}^\dagger K^\pm_i=\frac{1}{2}I_4,\ \ \ \forall\ i.$ This essential property is a result of the multiplication relations 
between $J_{mn}$ with equal and distinct indices and the 
fact that the set $\{1,2,3,4\}$ can be paritioned into three distinct sets of pairs of indices  $$\{(1,2),\ (3,4)\}, \ \  \{(1,3),\ (2,4)\}, \ \ \  \{(1,4),\ (2,3)\} $$
in such a way that in each partition every label appears only once, and the three partitions exhaust all the possible pairs. Such partions exist in even dimensions and their construction can be related to other interesting combinatorial problems in graph theory, scheduling and Latin squares. Note also that the new set of Kraus operators is obtained from the original set by the following transformation
\be
\begin{pmatrix}K^+_1\\ K^-_1\\ K^+_2\\ K^-_2\\K^+_3\\ K^-_3 \end{pmatrix}=\Omega\begin{pmatrix}J_{12}\\ J_{34}\\ J_{13}\\ J_{24}\\ J_{14}\\ J_{23} \end{pmatrix}
\ee
where $\Omega=H\oplus H\oplus H$ and $H=\frac{1}{\sqrt{2}}\begin{pmatrix}1& 1\\ 1& -1 \end{pmatrix}$ is a unitary matrix. This guarantees that the new Kraus operators define the same quantum channel as the original one.  Therefore by defining the unitary operators $U_i^\pm :=\sqrt{2}K_i^\pm$, the SO(4) LS channel can be written as 
\be
\Phi (\rho)=\frac{1}{6}\sum_{i=1}^3\big(U^+_i\rho {U^+_i}^\dagger+U^-_i\rho {U^-_i}^\dagger\big).
\ee
This shows that the channel simply acts as a mixture of unitary channels. This is also true for the channel $\Phi_x$, where one of the unitary operators is the identity operator. 
\subsection{A mixed unitary representation for $SO(6)$ Landau-Streater channel}
There are now in total 15 Kraus operators for this channel in the form
$J_{mn},\ 1\leq m<n\leq 6$. We now consider the following partition of indices, 
\ba
&& \{(1,2),\ (3,6),\ (4,5)\}\cr
&& \{(1,3),\ (2,4),\ (5,6)\}\cr
&&  \{(1,4),\ (3,5),\ (2,6)\}\cr
&&  \{(1,5),\ (2,3),\ (4,6)\}\cr
&&  \{(1,6),\ (2,5),\ (3,4)\}.
\ea
which have the nice property that in each set, each of the labels appear only once, while all the partitions are mutually exclusive, and exhaust all the pair of labels. Corresponding to each partition, say the first one, one can construct three unitary Kraus operators as follows: 
\be\label{Komega}
K_1=\frac{1}{\sqrt{3}}\big[J_{12}+J_{36}+J_{45}\big],\ \ \ \ K_2=\frac{1}{\sqrt{3}}\big[J_{12}+\omega J_{36}+\omega^2 J_{45}],\ \ \ \ K_3=\frac{1}{\sqrt{3}}\big[J_{12}+\omega^2 J_{36}+\omega J_{45}]
\ee
These operators satisfy $K_i^\dagger K_i=\frac{1}{3}I, \ \ \ \forall i$. In order to represent the channel as a mixture of unitary channels, we construct new Kraus operators out of the old ones in the same way as in (\ref{Komega}) for each of the paritions. Thus we have in a compact form
\be
{\bf K}=\Omega {\bf J}
\ee
where ${\bf K}$ is a column vector which comprises all the 15 new unitary Kraus operators, ${\bf J}$ is a vector which comprises all the original Kraus operators (in suitable order, i.e. corresponding to the consecutive partitions, like 
$${\bf J}=\begin{pmatrix}J_{12}& J_{36}& J_{45}& J_{13}& J_{24}& J_{56}&\cdots&\cdots\end{pmatrix}^T$$ and $\Omega = F\oplus F\oplus F\oplus F\oplus F$, in which $F=\frac{1}{\sqrt{3}}\begin{pmatrix}1 &1&1\\ 1&\omega &\omega^2\\ 1 & \omega^2&\omega\end{pmatrix}$ is the Fourier transform on $Z_3$. In this way and by defining $U_i=\sqrt{3}K_i$, the $SO(6)$ Landau-Streater channel is written as a uniform mixture of unitary maps:
\be
\Phi(\rho)=\frac{1}{15}\sum_i U_i \rho U_i^\dagger
\ee 
\\

\begin{figure}[H]
	\centering
	
	\includegraphics[width=\textwidth]{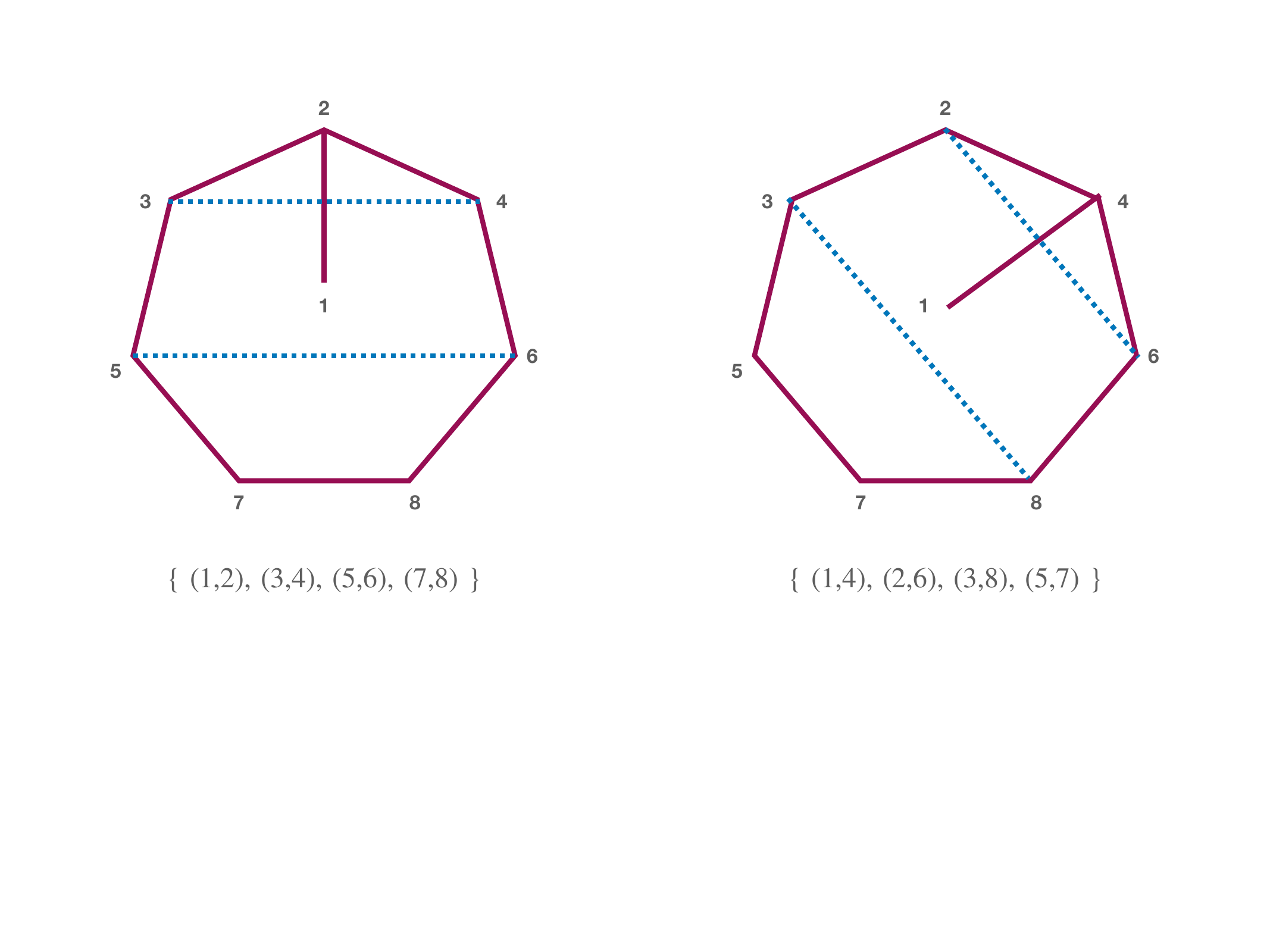}\vspace{-4cm}
	
	\caption{The algorithm for finding the distinct partitions of $d$ points for $d$=even. A point is put at the center and a line connects it with a point on the $d-1$ gon. The other pairs correspond to the lines perpendicular to this line. Here two examples are shown for $d=8$.  }
	\label{partition}
\end{figure}

\noindent The pattern displayed in these two examples repeats in higher dimensional channels provided that $d$ is even. In such cases there are $d-1$ different and mutually exclusive partitions. When $d$ is even, this kind of partitioning  corresponds to coloring the nodes of a graph with $d-1$ different colors in such a way that each node is connected to $d-1$ other nodes with  different colors. This problem is well known to have an algorithmic solution as shown in figure (\ref{partition}).

Each partition $I=\{(m,n),\cdots\}$ contains $d/2$ pairs of indices which allows us to convert the Kraus operators $\{J_{mn},\cdots\}$ by  (a Fourier Transform) to $d/2$ unitary Kraus operators. In this way the total number of $\frac{d(d-1)}{2}$ Kraus operators, are converted to the same number of unitary operators, describing the same quantum channel.     
For $d$ odd, the required partitions cannot be found  and it is not clear if the channel can admit a mixed unitary representation.

\section{Capacities}\label{classicalcapacity}

	For a quantum channel, one can define many different capacities \cite{gyongyosi_2018}\cite{arqand_quantum_2020}. These are the ultimate rates at which classical or quantum information can be transferred from a sender to a receiver per use of the channel by using different kinds of resources. There is a long route for converting these operational definitions to concrete and closed formulas for the capacities. Here we do not start from the operational definition for which the reader can refer to many good reviews \cite{gyongyosi_2018}\cite{wilde_book_2017}\cite{lloyd_capacity_1997}, rather we start from the closed formulae which have been obtained for the calculation of capacity in each case \cite{schumacher1997sending}\cite{barnum_information_1998}\cite{devetak_capacity_2005}\cite{devetak_private_2005}. Even after having these closed formulas, it is in general very difficult to find explicit values for the capacities in terms of the parameters of the channel and we have to suffice with bounds on these capacities \cite{Fanizza_2020},\cite{Kianvash_2022},\cite{Hirche_2023}. 
		Besides super-additivity ,\cite{Leditzky_2023}, the important property whose presence (or absence),  simplifies(or not) the calculation of some of these capacities  is the concavity of the relevant quantity which is to be maximized. We will see this in the following subsections, where we discuss different forms of capacities for the SO(d) Landau Streater channel.  

\subsection{One shot classical capacity}
This is the ultimate rate that  classical messages, when encoded into quantum states, can be transmitted reliably over a channel.  It is given by  \cite{schumacher1997sending}: 
\begin{equation} \label{classicalcap}
	\begin{split}
		C({\cal E})= \lim_{n\longrightarrow \infty} \frac{1}{n} \chi({\cal E}^{\otimes n}),
	\end{split}     
\end{equation}
where $\chi({\cal E})= \max_{p_i , \rho_i} \ \chi\{p_i,{\cal E}(\rho_i)\}$ \cite{wilde_book_2017} and $\chi\{p_i,\rho_i\}$ is the Holevo quantity of the output ensemble of states $\{p_i,\rho_i\}$ which is  defined as
\be
\chi(\{p_i,(\rho_i)\}):=S(\sum_i p_i\rho_i)-\sum_i p_iS(\rho_i).
\ee
Here $S(\rho)=-Tr(\rho\log \rho)$ is the von-Neumann entropy \cite{nielsen}. 
In general $\chi$ is superadditive, meaning that $n\chi({\cal E})\leq \chi({\cal E}^{\otimes n}) $, which makes the regularization in equation (\ref{classicalcap}) necessary for the calculation of the capacity \cite{hastings_superadditivity_2009}. This regularization is almost impossible so we suffice to calculate the one-shot classical capacity which is a lower bound for the full classical capacity.  It has been shown in \cite{wilde_book_2017} that one can only maximize $\chi\{p_i,{\cal E}(\rho_i)\}$   over ensembles of pure input states.
 As expected the covariance properties of the channel plays a significant role in the analytical form of this capacity. 
Let the minimum output entropy state be given by $|\psi\ra$. Now that we have lost the $U(d)$ covariance, we cannot transform this state $|\psi\ra$  into a given reference state of our choice. Instead we follow a different route and see how far we can proceed by reducing the parameters of the state $|\psi\ra$, by exploiting the $O(d)$ covariance. Let $|\psi\ra$ be of the form  
\be
|\psi\ra=\begin{pmatrix}\psi_1\\ \psi_2 \\ \psi_3 \\ \cdot \\ \psi_d\end{pmatrix}
\ee
where modulo a global phase, all the parameters $\psi_i$ are complex numbers and subject to normalization. We first use a rotation, generated by $J_{12}$, to transform 
$ \psi_2\lo -\sin\theta\  \psi_1+\cos\theta \ \psi_2$
to remove the imaginary part of $\psi_2$ and make it real, denoted hereafter by $r_2$. By successively using covariance generated by $J_{13}$, $J_{14},\cdots J_{1d}$, we make all the other coefficients $\psi_3,\psi_4,\cdots \psi_d$ real, making $|\psi\ra$ of the form
\be
|\psi\ra=\begin{pmatrix}\psi_1\\ r_2 \\ r_3 \\ \cdot \\ r_d\end{pmatrix}\h r_i\in R.
\ee
We are now ready to use rotations generated by $J_{23}, J_{24},\cdots J_{2d}$ to make all the parameters $r_i$ except $r_2$ to vanish, casting the state $|\psi_\ra$ into the form 
\be
\begin{pmatrix}\psi_1\\ r_2 \\ 0 \\ \cdot \\ 0\end{pmatrix}=\begin{pmatrix}\cos\theta e^{i\phi}\\ \sin\theta \\ 0 \\ \cdot \\ 0\end{pmatrix}
\ee
The output state will then be given by
\be
\begin{split}
	\Phi_{_{x}}(|\psi\ra\la \psi|)&=(1-x)\begin{pmatrix}\cos^2\theta & \cos\theta \sin\theta e^{i\phi}\\ \cos\theta \sin\theta e^{-i\phi}& \sin^2\theta\end{pmatrix}\oplus {\bf 0}^{d-2}\\
	&+(\frac{x}{d-1})\Big[\begin{pmatrix}1 & 0\\ 0& 1\end{pmatrix}\oplus  \mathcal{I}_{d-2}
	-\begin{pmatrix}\cos^2\theta & \cos\theta \sin\theta e^{-i\phi}\\ \cos\theta \sin\theta e^{i\phi}& \sin^2\theta\end{pmatrix}\oplus {\bf 0}^{d-2}\Big]
\end{split}
\ee
\ni This means that the eigenvalues of this output state comprise the disjoint union of two sets, namely:
\be\label{spectrumout}
{\rm Spectrum\ \  of }\big[ \Phi_{_{x}}(|\psi\ra\la \psi|)\big]=\{ \frac{x}{d-1} , g=d-2\} \cup {\rm Spectrum\ \  of }\ M
\ee
where as usual $g=d-2$ denotes the multiplicity of the first eigenvalue and $M$ is a two-dimensional matrix 
\be
M=\begin{pmatrix}(1-x)\cos^2\theta + \frac{x}{d-1}\sin^2\theta& -B\cos\theta \sin\theta \\ -B^*\cos\theta \sin\theta& (1-x)\sin^2\theta + \frac{x}{d-1}\cos^2\theta\end{pmatrix}.
\ee
Here $B$ is equal to 
\be
B=(1-x) e^{i\phi}+\frac{x}{d-1}e^{-i\phi}.
\ee
In order to find the minimum output entropy state, we do not need to explicitly find the eigenvalues of this matrix. It suffices to note that the trace of this matrix, which is the sum of its eigenvalues is equal to 
\be\label{sumeig}
\tr M= \lambda_1+\lambda_2=1-x+\frac{x}{d-1},
\ee
and is independent of the input state, while its determinant which is the product of its eigenvalues is equal to 
\be
\det(M)=\frac{x(1-x)}{d-1}\Big[\cos^4 \theta + \sin^4\theta-2\cos^2\theta\sin^2\theta \cos 2\phi\Big]
\ee

\noindent Since  $\lambda_1+\lambda_2$ is independent of the input state, the entropy is minimized if we minimize $\lambda_1\lambda_2$ or the determinant of $M$. We can minimize $Det(M)$ by setting $\phi=\frac{\pi}{2}$ and $\theta=\frac{\pi}{4}$ which leads to a vanishing $\det(M)$.  The minimum output entropy state is now of the form
	\be
\theta=\frac{\pi}{4}\lo |\psi\ra=\frac{1}{\sqrt{2}}\begin{pmatrix}i \\ 1 \\ 0 \\ \cdot \\ 0\end{pmatrix}.
\ee
		In view of (\ref{spectrumout}), the complete set of eigenvalues of $\Phi_x(|\psi\ra\la \psi)$ will now be given by 
		\be
		{\rm Spectrum\ \  of }\big[ \Phi_{_{x}}(|\psi\ra\la \psi|)\big]=\{ \frac{x}{d-1} , g=d-2\} \cup \{0, 1-x+\frac{x}{d-1}\}.
		\ee
This leads to the following value for the classical one-shot capacity		
		\be\label{capB}
		C^1(\Phi_{_{x}})=\log d+(d-2)\frac{x}{d-1}\log(\frac{x}{d-1})+(1-x+\frac{x}{d-1})\log(1-x+\frac{x}{d-1}).
		\ee
This capacity interpolates between $\log d$ for the identity channel $\Phi_0$ and $\log d-\log (d-1)$ for the Werner-Holevo channel $\Phi_1$. The value for the WH channel can be intuitively understood if we note that any input state of the form $|\psi\ra=|i\ra$ sent by Alice is received by Bob as $\frac{1}{d-1}(I-|i\ra\la i|)$. This leads to the following conditional probabilities 
$$P(y=j|x=i)=\frac{1}{d-1}(1-\delta_{ij}).$$ With $P(x_i)=\frac{1}{d}$, this leads to $P(x=i,y=j)=\frac{1}{d(d-1)}(1-\delta_{ij})$ and $P(y=j)=\frac{1}{d}$, leading to the following value for mutual quantum information
\be
I(X:Y)=\log d+\log d+\frac{1}{d(d-1)}\sum_{i,j}(1-\delta_{ij})\log(\frac{1-\delta_{ij}}{d(d-1)}) =\log d-\log (d-1).
\ee

\subsection{Entanglement-Assisted Classical Capacity}\label{ent}
Entanglement-assisted capacity is a measure of the maximum rate at which quantum information can be transmitted through a noisy quantum channel when the sender and receiver are allowed to share an unlimited number of entangled quantum state \cite{bennett_entanglement-assisted_1999}. The entanglement-assisted classical capacity of a given channel $\Lambda$ is determined by \cite{bennett_entanglement-assisted_2002}:
\begin{equation}
	C_{ea}({\cal E})=\max_{\rho} I(\rho,\Phi)
\end{equation}
where 
\be
I(\rho, {\cal E}):=S(\rho)+S( {\cal E}(\rho))-S(\rho, {\cal E}).
\ee
Here $S(\rho,{\cal E})$ is the output entropy of the environment, referred to as the entropy exchange \cite{nielsen_1998}, and is represented by the expression $S(\rho,{\cal E}) = S({\cal E}^c(\rho))$ where ${\cal E}^c$ is the complementary channel \cite{HolevoComp}.
According to proposition 9.3 in \cite{holevoBook}, the maximum entanglement-assisted capacity of a covariant channel  is attained for an invariant state $\rho$. In the special case where is irreducibly covariant, the maximum is attained on the maximally mixed state.
Hence, for the channel $\Phi_{_{x}}$, we have 
\begin{equation}
	C_{ea}(\Phi_{_{x}})=S(\frac{I}{d})+S(\Phi_{_{x}}(\frac{I}{d}))-S(\Phi^c_{_{x}}(\frac{I}{d})),
\end{equation}
which, given the unitality of the channel, leads to
\begin{equation}\label{Cee}
	C_{ea}(\Phi_{_{x}})=2\log_2d-S(\Phi^c_{_{x}}(\frac{I}{d})).
\end{equation}
From (\ref{compblock}), we find 
\be
\Phi_{_{x}}^c(\frac{I}{d})=\begin{pmatrix} (1-x) & {\bf 0}^T\\
	{\bf 0}&\frac{x}{d(d-1)}(I\otimes I-S)\end{pmatrix}.\ee
This matrix is of dimension $(1+d^2)\times (1+d^2)$. The lower corner is the tensor product of $d$ dimensional square matrices.  
With the notation $|m,n\ra:=|m\ra\otimes |n\ra\in \mathcal{H}_{d^2}$, 
an  eigenvector of this matrix is given by $\begin{pmatrix}  1\\ {\bf 0}\end{pmatrix}\in \mathcal{H}_1\oplus \mathcal{H}_{d^2}$, corresponding to eigenvalue $(1-x)$. There are also $\frac{d(d+1)}{2}$ eigenvectors of the form $\begin{pmatrix}  0\\ |m,n\ra+|n,m\ra\end{pmatrix}\in  \mathcal{H}_1\oplus \mathcal{H}_{d^2}$ with vanishing eigenvalues and $\frac{d(d-1)}{2}$ eigenvectors of the form $\begin{pmatrix}  0\\ |m,n\ra-|n,m\ra\end{pmatrix}\in  \mathcal{H}_1\oplus \mathcal{H}_{d^2}$ with  eigenvalues equal to $\frac{2x}{d(d-1)}$. Hence the entanglement-assisted capacity is equal to   
\begin{equation}\label{Ceer}
	C_{ea}(\Phi_{_{x}})=2\log_2d+(1-x)\log_2(1-x)+x\log_2\frac{2x}{d(d-1)}.
\end{equation}
This interpolates between $2\log_2 d$ for the identity channel $\Phi_0$ (as it should be) and $1+\log_2d-\log_2(d-1)$ for the pure Landau-Streater channel, which is one bit larger than the one-shot classical capacity for the LS channel. When $d=2$, the Werner-Holevo or the Landau-Streater channel has only one Kraus operator given by $\sigma_y=\begin{pmatrix}0&-i\\ i&0\end{pmatrix}$, meaning that the channel acts in a unitary way. In this case the classical capacity is equal to one-bit and the entanglement-assisted capacity is equal to 2-bits per use of the channel which are what we expect from the dense-coding protocol. 

\subsection{ Bounds for  the quantum capacity}
This section is a modest attempt to find a lower bound for the quantum capacity in the form of the one-shot quantum capacity $Q^1(\Phi_{_{x}})$. It can only serve as a starting point for more detailed investigation of this problem. \\

\noindent Given a quantum channel ${\cal E}$, the quantum capacity $Q({\cal E})$ 
is the ultimate rate for transmitting quantum information and preserving the entanglement between the channel's input and a reference quantum state over a quantum channel.  This quantity is described in terms of coherent information \cite{barnum_information_1998, lloyd_capacity_1997,shor_quantum_2002,devetak_private_2005}:
\begin{equation}
	\begin{split}
		Q({\cal E}) =
		\lim_{n \to \infty}  \; \frac{1}{n} J(
		{\cal E}^{\otimes n}),
	\end{split}     
\end{equation}
where $J({\cal E})=\max_{\rho} J(\rho,{\cal E})$ and $J(\rho,{\cal E}):=S({\cal E}(\rho))-S({\cal E}^c(\rho))$. It is known that $J$  is superadditive, i.e. $J({\cal E}_1 \otimes {\cal E}_2)\ge J({\cal E}_1)+J({\cal E}_2)$, rendering an exact calculation of the quantum capacity extremely difficult and at the same time providing a lower bound in the form  $Q^{(1)}(\Lambda)\leq Q({\cal E})$ where $Q^{(1)}:=J({\cal E})$ is the single shot capacity.  However if the channel is  degradable, then the additivity property is restored $Q({\cal E})=Q^{(1)}({\cal E})$ \cite{devetak_capacity_2005}, and the calculation of the quantum capacity becomes a convex optimization problem. Approximate degradability, as defined and investigated in \cite{Sutter_2017}, can provide lower and upper bounds for the quantum capacity. For the modified $SO(3)$ Landau-Streater channel, approximate degradibility has been recently investigated in \cite{lo2024degradability}. For the so(d) Landau-Streater channel, we do not address this problem and instead suffice to determine a lower bound for the quantum capacity in the form of single-shot quantum capacity.  We therefore start with  
\be
Q^{(1)}(\Phi_{_{x}})=Max_{\rho}J(\rho,\Phi_{_{x}})=Max_{\rho}\Big[S(\Phi_{_{x}}(\rho))-S(\Phi_{_{x}}^c(\rho))\Big]
\ee
This shows that  any state $\rho$, not the one which maximizes the coherent information will also provide a lower bound for the quantum capacity although it may not a tight bound. 
{\it We stress that the lower bound that we find is by no means a tight lower bound. It is just a starting point for more detailed investigation of this problem. } Let us restrict our search within the real density matrices. Since both $\Phi_{_{x}}$ and $\Phi_{_{x}}^c$ are covariant under $SO(d)$ transformations, we can safely use such a transformation to diagonalize $\rho$ and put it in the form 
\be
\rho_0\equiv U\rho U^\dagger=\begin{pmatrix}r_1&&&&\\ &r_2&&&\\ \cdot&\cdot&\cdot&\cdot&\cdot \\\cdot&\cdot&\cdot&\cdot&\cdot\\ &&&& r_d\end{pmatrix}
\ee
where
\be
\sum_{i=1}^dr_i=1
\ee
The action of the channel $\Phi_{_{x}}$ on $\rho_0$ is directly found from the definition of the channel in (\ref{deff}). The result is a diagaonal matrix $\Phi_{_{x}}(\rho_0)=Diag(\lambda_1,\lambda_2,\cdots \lambda_d)$, so that 
\be
S(\Phi_{_{x}}(\rho_0))=-\sum_{i=1}^d \lambda_i\log_2\lambda_i
\ee
where
\be\label{lambdai}
\lambda_i=(1-x)r_i+\frac{x}{d-1}(1-r_i).
\ee
We also have  to find the spectrum of the matrix $\Phi^c_{_{x}}(\rho_0)$. When we write the channel in the form 
(\ref{symmcomplement}), the complement channel has $d^2+1$ Kraus operators and hence this square matrix is of dimension $d^2+1$ as given in (\ref{compblock}). As argued after that equation, when $\rho$ is a diagonal matrix, the off-diagonal blocks vanish and we are left with
\be\label{compdiagonal}
\Phi_{_{x}}^c(\rho)=\begin{pmatrix} (1-x)\tr(\rho) & {\bf 0}^T\\ 
	{\bf 0}&\frac{x}{2(d-1)}(I-S)(I\otimes \rho + \rho\otimes I)\end{pmatrix}.\ee
The form (\ref{symm}) shows that for any diagonal matrix the off-diagonal blocks vanish and we are left with
\be\label{compdiagonalfinal}
\Phi_{_{x}}^c(\rho)=\begin{pmatrix} (1-x)\tr(\rho) & {\bf 0}^T\\ 
	{\bf 0}&\frac{x}{2(d-1)}D\end{pmatrix}.\ee
where \ba D&=&(I-S)(I\otimes \rho + \rho\otimes I)=(I-S)\sum_{i,j}\sum_{i,j}(r_i+r_j)|i,j\ra\la i,j|\cr
&=&\sum_{i,j}(r_i+r_j)(|i,j\ra-|ji\ra)\la i,j|=\sum_{i,j}(r_i+r_j)|e_{ij}\ra\la e_{ij}|=2\sum_{i<j}(r_i+r_j)|e_{ij}\ra\la e_{ij}|
\ea
where $|e_{ij}\ra:=\frac{1}{\sqrt{2}}(|i,j\ra-|j,i\ra)$. 
With diagonalizaiton of $D$, the full spectrum of $\Phi_x^c(\rho)$ is determined. Combining all these, one finds 
\be\begin{split}
	J(\rho, \Phi_{_{x}})&=-\sum_{i=1}^d \lambda_i\log_2  \lambda_i+(1-x)\log_2(1-x) +  \sum_{i<j}\frac{x(r_i+r_j)}{d-1}\log_2\frac{x(r_i+r_j)}{d-1},
\end{split}
\ee
where $\lambda_i$ is given in (\ref{lambdai}). 
\noindent One can obtain various lower bounds by taking simple density matrices like  $$\rho_n=Diagonal(\frac{1}{n},\frac{1}{n},\cdots,\frac{1}{n},0,0,\cdots 0).$$  Insertion of this density matrix in the above formula, leads to 
\ba\label{Jn}
J(\rho_n,\Phi^d_x)&=&-[(1-x)+x\frac{n-1}{d-1}]\log(\frac{1-x}{n}+\frac{x(n-1)}{n(d-1)})\cr &+&(\frac{n-1}{d-1})x\log(\frac{x}{d-1})+(1-x)\log(1-x)-x\log(n)+\frac{x(n-1)}{d-1}
\ea
where for clarity we have temporarily added a superscipt $d$ to the notation of the channel. This function has several interesting properties: \\

{\bf i-} When $x=0$ and we are dealing with the identity channel, it is evident that $J(\rho_n,\Phi^d_{x=0})=\log n$ where its maximum is achieved for the completely mixed state, i.e. for $n=d$. This gives a lower bound of $Q^1(\Phi^d_0)=\log d$, which is in fact equal to the quantum capacity of the idenity channel. \\

\begin{figure}[H]
	\centering
	
	\includegraphics[width=8 cm]{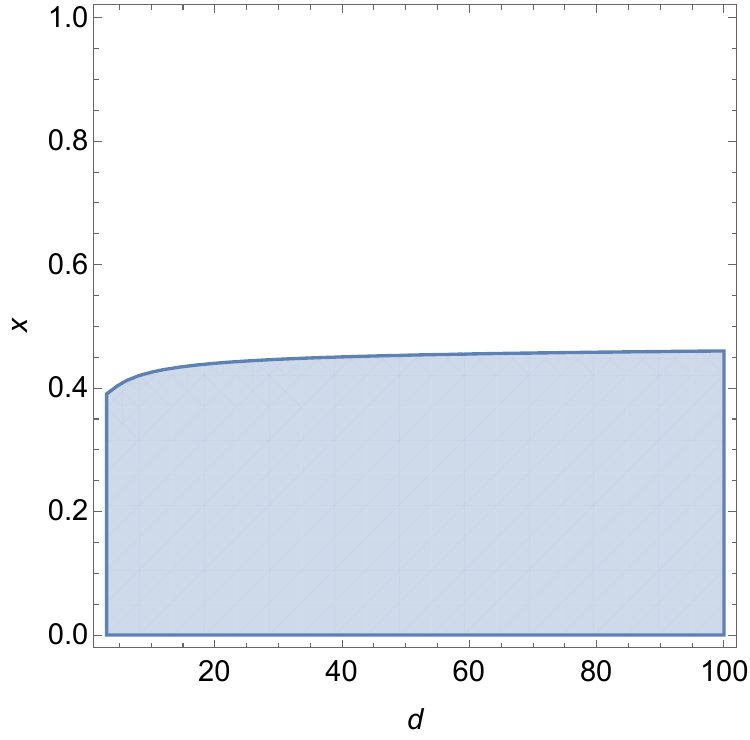}\vspace{0.cm}
	
	\caption{The shaded region shows the range of the parameter $x$ in which the channel $\Phi_x$ has a positive quantum capacity. Interestingly the cricial value is approximately equal to $0.4$ for all dimensions. Above this critical value, the coherent information is negative and it is not known if the channel $\Phi_{x>x_0}$ has a non-zero quantum capacity.  }
	\label{criticalX}
\end{figure}

{\bf ii-} For a pure input state $\rho_1$, we see that 
$J(\rho_1,\Phi^d(x))=0,\ \ \forall\ d\ {\rm and}\ x.$ This does not give any useful lower bound for the quantum capacity. However for a maximally mixed state $\rho_d$, we find from (\ref{Jn}) that 
\be
J(\rho_d,\Phi_x)=(1-x)\log_2 d(1-x)+x(1+\log_2\frac{x}{d-1}),
\ee
which can be positive if the parameter $x$ is less than a certain critical value  $x_{0}$. This indicates that the channel $\Phi_{x<x_0}$ will have a positive quantum capacity. Numerical solution of  $$(1-x)\log_2 d(1-x)+x(1+\log_2\frac{x}{d-1})\geq 0,$$ determines this critical value. Figure (\ref{criticalX}) show interestingly that for all dimensions $d$, $x_0\approx 0.4$. This is in accord with the result of \cite{roofeh2023noisy} where semidifinite programming was used for the case of $d=3$ (the modified so(3) Landau-streater channel). 

\section{The Landau-Streater channel for the group $SU(d)$}
In this section, we briefly mention how the LS channel can be generalized to the group $U(d)$. 
Let us  replace the set $\Delta_-=\{J_{m,n}\}$ with the following set of operators $\Delta_+:=\{K_{mn},1\leq  m, n\leq d\}$
where 
\be\label{KK}
K_{mn}=|m\ra\la n|+|n\ra\la m|,\ \ \  m<n, \h K_{mm}=2|m\ra\la m|.
\ee		
The set $\Delta_+$ of dimension $\frac{d(d+1)}{2}$ is not closed under the Lie-bracket and hence is not a Lie-algebra anymore. However the combined set $\Delta=\Delta_-\cup \Delta_+$ forms the Lie-Algebra of the group $U(d)$ of $d$-dimensional unitary matrices. Note that dimension of $\Delta$, i.e. is equal to $d^2$ which is equal to the dimension of the group $U(d)$. 
One can show by direct calculation that 
\be
\frac{1}{2}\sum_{m, n}K_{mn}^\dagger K_{mn}=(d+1)\ \mathcal{I}.
\ee	
Therefore one can define another quantum channel as follows
\be
\Phi^+(\rho):=\frac{1}{2(d+1)}\sum_{m,  n}K_{mn}\rho K_{mn}^\dagger = \frac{1}{d+1}\left[\tr\rho\  \mathcal{I}+\rho^T)\right]
\ee
This is a new generalization of the Landau-Streater channel which is equivalent to the Werner-Holevo channel $\Phi_{_{1,d}}$ (for the notation see below). Even if the Kraus opeators $K_{mn}$ are not generators of a Lie algebra anymore. Despite this, the covariance under $U(d)$ holds for both channels, that is
\be
\Phi^{\pm}(U\rho U^\dagger)=U^*\Phi^{\pm}(\rho)U^T\h U\in U(d)
\ee
where $U(d)$ is the group of unitary operators on $\mathcal{H}_d$.
We can now make the following convex combination of these channels to arrive at a one-parameter family of channels:
\be
\Phi_{{\eta,d}}(\rho)=\frac{1-\eta}{2}\Phi^-(\rho) + \frac{1+\eta}{2} \Phi^+(\rho)
\ee
which turns out to be equal to the one-parameter family of Werner-Holevo channels in equation (\ref{whcope}) \cite{WH}. The Kraus operators are now the set of generators of the Lie-Algebra $u(d)$, namely the set $\{ J_{mn},\ K_{mn}\}$. Therefore we call this the U(d)  Landau-Streater channel. One also modify this channel to represent a two-parameter family of noisy su(d) LS-channels by defining a convex combination of the identity channel, $\Phi^-$ and $\Phi^+$.

\section{Discussion}
We have generalized and studied in rather great detail, the Landau-Streater channel which is pertaining to the spin-$j$ representation of the Lie-algebra of the group $SO(3)$ to the fundamental representation of the groups $SO(d)$ and  have pointed out their equivalence to the Werner-Holevo channels. We have studied the so-called noisy versions of these channels and have determined several properties of the resulting one-parameter family of quantum channels, including their spectrum, their region of infinitesimal divisibility, their complement channels and finally their one-shot classical capacity and their entanglement-assisted classical capacity. We have also found a lower bound for the quantum capacity of this modified channel and have shown that in all dimensions, when the noise parameter is less than a critical value approximately equal to $0.4$, the channel has a non-zero quantum capacity. While the pure $SO(3)$ Landau-Streater channel is known to be an extreme point in the space of channels, we have shown that  the pure $SO(d)$ channel is not extreme. Moreover we have found a mixed unitary representation for it, when $d$ is an even number. \\

 It would be an interesting problem if one can define and study the Landau-Streater channel in a most general setting, namely any representation of any Lie-algebra \cite{Ritter_2005}.  
 Certainly these kinds of channels may not find concrete applications in quantum information processing, but they are definitely of great interest in the structural theory of quantum channels and completely positive maps. As pointed out in \cite{lo2024degradability}, " understanding these channels is critical in shedding light on the super-additivity effect in quantum channels operating in low-noise regimes" as the  "the long-term goal is to extend this desirable property to a wider spectrum of quantum channels beyond the afore- mentioned generalizations of the qubit depolarizing channel, thereby enriching our understanding of the super-additivity effect in high-dimensional low-noise scenarios" .\\
 
  Even for the $SO(d)$ groups and the representation we have used, our study can lead to many extensions, the immediate one will be to study their approximate degradability along the lines of the recent work \cite{lo2024degradability}.  Another one is to study a two-parameteric family of extensions of the Landau-Streater channel for the group $SU(d)$. This will be the subject of a future work. 

	\section{Acknowledgements} 
 This research was supported in part by Iran National Science Foundation, under Grant No.4022322. I would like to thank members of the QIS group in Sharif, especially Shayan Roofeh for the algorithmic solution leading to figure (\ref{partition}) and Abolfazl Farmanian for their valuable comments.  I also thank Farzad Kianvash and Laleh Memarzadeh for valuable discussions.

\newpage
\bibliography{refs}

\begin{thebibliography}{10}

\bibitem{Datta_2005}
Nilanjana Datta and Mary~Beth Ruskai.
\newblock Maximal output purity and capacity for asymmetric unital qudit
  channels.
\newblock {\em Journal of Physics A: Mathematical and General},
  38(45):9785–9802, October 2005.

\bibitem{smith2010quantum}
Graeme Smith.
\newblock Quantum channel capacities, 2010.

\bibitem{shor_2003}
Peter~W. Shor.
\newblock Capacities of quantum channels and how to find them.
\newblock {\em Mathematical Programming}, 97(1):311–335, July 2003.

\bibitem{Chessa_2021}
Stefano Chessa and Vittorio Giovannetti.
\newblock Quantum capacity analysis of multi-level amplitude damping channels.
\newblock {\em Communications Physics}, 4(1), February 2021.

\bibitem{Islam_2017}
Nurul~T. Islam, Charles Ci~Wen Lim, Clinton Cahall, Jungsang Kim, and Daniel~J.
  Gauthier.
\newblock Provably secure and high-rate quantum key distribution with time-bin
  qudits.
\newblock {\em Science Advances}, 3(11), November 2017.

\bibitem{Karimipour_2002}
Vahid Karimipour, Alireza Bahraminasab, and Saber Bagherinezhad.
\newblock Quantum key distribution ford-level systems with generalized bell
  states.
\newblock {\em Physical Review A}, 65(5), May 2002.

\bibitem{Wang_2020}
Yuchen Wang, Zixuan Hu, Barry~C. Sanders, and Sabre Kais.
\newblock Qudits and high-dimensional quantum computing.
\newblock {\em Frontiers in Physics}, 8, November 2020.

\bibitem{Rubinsztein-Dunlop_2017}
Halina Rubinsztein-Dunlop, Andrew Forbes, M~V Berry, M~R Dennis, David~L
  Andrews, Masud Mansuripur, Cornelia Denz, Christina Alpmann, Peter Banzer,
  Thomas Bauer, Ebrahim Karimi, Lorenzo Marrucci, Miles Padgett, Monika
  Ritsch-Marte, Natalia~M Litchinitser, Nicholas~P Bigelow, C~Rosales-Guzmán,
  A~Belmonte, J~P Torres, Tyler~W Neely, Mark Baker, Reuven Gordon, Alexander~B
  Stilgoe, Jacquiline Romero, Andrew~G White, Robert Fickler, Alan~E Willner,
  Guodong Xie, Benjamin McMorran, and Andrew~M Weiner.
\newblock Roadmap on structured light.
\newblock {\em Journal of Optics}, 19(1):013001, nov 2016.

\bibitem{Gao_2024}
Xiaoqin Gao, Yingwen Zhang, Alessio D’Errico, Alicia Sit, Khabat Heshami, and
  Ebrahim Karimi.
\newblock Full spatial characterization of entangled structured photons.
\newblock {\em Physical Review Letters}, 132(6), February 2024.

\bibitem{Karimi:12}
Ebrahim Karimi, Lorenzo Marrucci, Corrado de~Lisio, and Enrico Santamato.
\newblock Time-division multiplexing of the orbital angular momentum of light.
\newblock {\em Opt. Lett.}, 37(2):127--129, Jan 2012.

\bibitem{PhysRevApplied.11.064058}
Daniele Cozzolino, Davide Bacco, Beatrice Da~Lio, Kasper Ingerslev, Yunhong
  Ding, Kjeld Dalgaard, Poul Kristensen, Michael Galili, Karsten Rottwitt,
  Siddharth Ramachandran, and Leif~Katsuo Oxenl\o{}we.
\newblock Orbital angular momentum states enabling fiber-based high-dimensional
  quantum communication.
\newblock {\em Phys. Rev. Appl.}, 11:064058, Jun 2019.

\bibitem{Siudzinska_2020}
Katarzyna Siudzińska.
\newblock Classical capacity of generalized pauli channels.
\newblock {\em Journal of Physics A: Mathematical and Theoretical},
  53(44):445301, oct 2020.

\bibitem{Akio_Fujiwara_2003}
Akio Fujiwara and Hiroshi Imai.
\newblock Quantum parameter estimation of a generalized pauli channel.
\newblock {\em Journal of Physics A: Mathematical and General}, 36(29):8093,
  jul 2003.

\bibitem{WH}
R.~F. Werner and A.~S. Holevo.
\newblock Counterexample to an additivity conjecture for output purity of
  quantum channels.
\newblock {\em Journal of Mathematical Physics}, 43:4353, 2002.

\bibitem{LanS}
L.~J. Landau and R.~F. Streater.
\newblock On birkhoff's theorem for doubly stochastic completely positive maps
  of matrix algebras.
\newblock {\em Linear Algebra and its Applications}, 193:107, 1993.

\bibitem{Audenaert_2008}
Koenraad M~R Audenaert and Stefan Scheel.
\newblock On random unitary channels.
\newblock {\em New Journal of Physics}, 10(2):023011, February 2008.

\bibitem{filippov_LS}
Sergey~N. Filippov and Ksenia~V. Kuzhamuratova.
\newblock Quantum informational properties of the {Landau}–{Streater}
  channel.
\newblock {\em Journal of Mathematical Physics}, 60(4):042202, April 2019.

\bibitem{pakhomchik_realization_2020}
A.~I. Pakhomchik, I.~Feshchenko, A.~Glatz, V.~M. Vinokur, A.~V. Lebedev, S.~N.
  Filippov, and G.~B. Lesovik.
\newblock Realization of the {Werner}–{Holevo} and {Landau}–{Streater}
  {Quantum} {Channels} for {Qutrits} on {Quantum} {Computers}.
\newblock {\em Journal of Russian Laser Research}, 41(1):40--53, January 2020.

\bibitem{Girard_2022}
Mark Girard, Debbie Leung, Jeremy Levick, Chi-Kwong Li, Vern Paulsen, Yiu~Tung
  Poon, and John Watrous.
\newblock On the mixed-unitary rank of quantum channels.
\newblock {\em Communications in Mathematical Physics}, 394(2):919–951, June
  2022.

\bibitem{datta2004additivity}
Nilanjana Datta, Alexander~S. Holevo, and Yuri Suhov.
\newblock Additivity for transpose depolarizing channels, 2004.

\bibitem{Cope_2017}
Thomas P.~W. Cope and Stefano Pirandola.
\newblock Adaptive estimation and discrimination of holevo-werner channels.
\newblock {\em Quantum Measurements and Quantum Metrology}, 4(1), December
  2017.

\bibitem{Cope_2018}
Thomas P~W Cope, Kenneth Goodenough, and Stefano Pirandola.
\newblock Converse bounds for quantum and private communication over
  holevo–werner channels.
\newblock {\em Journal of Physics A: Mathematical and Theoretical},
  51(49):494001, November 2018.

\bibitem{Chitambar_2023}
Eric Chitambar, Ian George, Brian Doolittle, and Marius Junge.
\newblock The communication value of a quantum channel.
\newblock {\em IEEE Transactions on Information Theory}, 69(3):1660–1679,
  March 2023.

\bibitem{Wolf_2005}
M~M Wolf and J~Eisert.
\newblock Classical information capacity of a class of quantum channels.
\newblock {\em New Journal of Physics}, 7:93–93, April 2005.

\bibitem{fannes2004additivity}
M.~Fannes, B.~Haegeman, M.~Mosonyi, and D.~Vanpeteghem.
\newblock Additivity of minimal entropy output for a class of covariant
  channels, 2004.

\bibitem{konrad}
KMR Audenaert and S~Scheel.
\newblock On random unitary channels.
\newblock {\em New Journal of Physics}, 10(2):023011, 2008.

\bibitem{Horodecki_2003}
Michael Horodecki, Peter~W. Shor, and Mary~Beth Ruskai.
\newblock Entanglement breaking channels.
\newblock {\em Reviews in Mathematical Physics}, 15(06):629–641, August 2003.

\bibitem{roofeh2023noisy}
Shayan Roofeh and Vahid Karimipour.
\newblock The noisy werner-holevo channel and its properties, 2023.

\bibitem{caves_qutrit_2000}
Carlton~M. Caves and Gerard~J. Milburn.
\newblock Qutrit entanglement.
\newblock {\em Optics Communications}, 179(1-6):439--446, May 2000.

\bibitem{brus_optimal_2002}
D.~Bruss and C.~Macchiavello.
\newblock Optimal eavesdropping in cryptography with three-dimensional quantum
  states.
\newblock {\em Physical Review Letters}, 88(12):127901, March 2002.

\bibitem{molina-terriza_experimental_2005}
G.~Molina-Terriza, A.~Vaziri, R.~Ursin, and A.~Zeilinger.
\newblock Experimental quantum coin tossing.
\newblock {\em Physical Review Letters}, 94(4):040501, January 2005.

\bibitem{kendon_bounds_2002}
Vivien~M. Kendon, Karol Życzkowski, and William~J. Munro.
\newblock Bounds on entanglement in qudit subsystems.
\newblock {\em Physical Review A}, 66(6):062310, December 2002.

\bibitem{cerf_greenberger-horne-zeilinger_2002}
Nicolas~J. Cerf, Serge Massar, and Stefano Pironio.
\newblock Greenberger-horne-zeilinger paradoxes for many qudits.
\newblock {\em Physical Review Letters}, 89(8):080402, August 2002.

\bibitem{bartlett_quantum_2002}
Stephen~D. Bartlett, Hubert de~Guise, and Barry~C. Sanders.
\newblock Quantum encodings in spin systems and harmonic oscillators.
\newblock {\em Physical Review A}, 65(5):052316, May 2002.

\bibitem{bouda_entanglement_2001}
Jan Bouda and Vladimír Buzek.
\newblock Entanglement swapping between multi-qudit systems.
\newblock {\em Journal of Physics A: Mathematical and General}, 34(20):4301,
  May 2001.

\bibitem{lo2024degradability}
Yun-Feng Lo, Yen-Chi Lee, and Min-Hsiu Hsieh.
\newblock Degradability of modified landau-streater type low-noise quantum
  channels in high dimensions, 2024.

\bibitem{devetak_capacity_2005}
I.~Devetak and P.~W. Shor.
\newblock The {Capacity} of a {Quantum} {Channel} for {Simultaneous}
  {Transmission} of {Classical} and {Quantum} {Information}.
\newblock {\em Communications in Mathematical Physics}, 256(2):287--303, June
  2005.

\bibitem{cubitt_structure_2008}
Toby~S. Cubitt, Mary~Beth Ruskai, and Graeme Smith.
\newblock The structure of degradable quantum channels.
\newblock {\em Journal of Mathematical Physics}, 49(10):102104, October 2008.

\bibitem{Sutter_2017}
David Sutter, Volkher~B. Scholz, Andreas Winter, and Renato Renner.
\newblock Approximate degradable quantum channels.
\newblock {\em IEEE Transactions on Information Theory}, 63(12):7832–7844,
  December 2017.

\bibitem{wolf_dividing_2008}
Michael~M. Wolf and J.~Ignacio Cirac.
\newblock Dividing {Quantum} {Channels}.
\newblock {\em Communications in Mathematical Physics}, 279(1):147--168, April
  2008.

\bibitem{stinespring}
W.~Forrest Stinespring.
\newblock Positive {Functions} on {C} * -{Algebras}.
\newblock {\em Proceedings of the American Mathematical Society}, 6(2):211,
  April 1955.

\bibitem{datta_complementarity_2006}
N.~Datta, M.~Fukuda, and A.~S. Holevo.
\newblock Complementarity and {Additivity} for {Covariant} {Channels}.
\newblock {\em Quantum Information Processing}, 5(3):179--207, June 2006.

\bibitem{smaczynski2016selfcomplementary}
Marek Smaczyński, Wojciech Roga, and Karol Życzkowski.
\newblock Selfcomplementary {Quantum} {Channels}.
\newblock {\em Open Systems \& Information Dynamics}, 23(03):1650014, September
  2016.

\bibitem{gyongyosi_2018}
Laszlo Gyongyosi, Sandor Imre, and Hung~Viet Nguyen.
\newblock A survey on quantum channel capacities.
\newblock {\em IEEE Communications Surveys amp; Tutorials}, 20(2):1149–1205,
  2018.

\bibitem{arqand_quantum_2020}
Amir Arqand, Laleh Memarzadeh, and Stefano Mancini.
\newblock Quantum capacity of a bosonic dephasing channel.
\newblock {\em Physical Review A}, 102(4):042413, October 2020.

\bibitem{wilde_book_2017}
In Mark~M. Wilde, editor, {\em Quantum {Information} {Theory}}, pages xi--xii.
  Cambridge University Press, Cambridge, 2 edition, 2017.

\bibitem{lloyd_capacity_1997}
Seth Lloyd.
\newblock Capacity of the noisy quantum channel.
\newblock {\em Physical Review A}, 55(3):1613--1622, March 1997.

\bibitem{schumacher1997sending}
Benjamin Schumacher and Michael~D. Westmoreland.
\newblock Sending classical information via noisy quantum channels.
\newblock {\em Physical Review A}, 56(1):131--138, July 1997.

\bibitem{barnum_information_1998}
Howard Barnum, M.~A. Nielsen, and Benjamin Schumacher.
\newblock Information transmission through a noisy quantum channel.
\newblock {\em Physical Review A}, 57(6):4153--4175, June 1998.

\bibitem{devetak_private_2005}
I.~Devetak.
\newblock The private classical capacity and quantum capacity of a quantum
  channel.
\newblock {\em IEEE Transactions on Information Theory}, 51(1):44--55, January
  2005.

\bibitem{Fanizza_2020}
Marco Fanizza, Farzad Kianvash, and Vittorio Giovannetti.
\newblock Quantum flags and new bounds on the quantum capacity of the
  depolarizing channel.
\newblock {\em Physical Review Letters}, 125(2), July 2020.

\bibitem{Kianvash_2022}
Farzad Kianvash, Marco Fanizza, and Vittorio Giovannetti.
\newblock Bounding the quantum capacity with flagged extensions.
\newblock {\em Quantum}, 6:647, February 2022.

\bibitem{Hirche_2023}
Christoph Hirche and Felix Leditzky.
\newblock Bounding quantum capacities via partial orders and complementarity.
\newblock {\em IEEE Transactions on Information Theory}, 69(1):283–297,
  January 2023.

\bibitem{Leditzky_2023}
Felix Leditzky, Debbie Leung, Vikesh Siddhu, Graeme Smith, and John~A. Smolin.
\newblock Generic nonadditivity of quantum capacity in simple channels.
\newblock {\em Physical Review Letters}, 130(20), May 2023.

\bibitem{nielsen}
Michael~A. Nielsen and Isaac~L. Chuang.
\newblock {\em Quantum computation and quantum information}.
\newblock Cambridge University Press, Cambridge ; New York, 10th anniversary ed
  edition, 2010.

\bibitem{hastings_superadditivity_2009}
M.~B. Hastings.
\newblock Superadditivity of communication capacity using entangled inputs.
\newblock {\em Nature Physics}, 5(4):255--257, April 2009.

\bibitem{bennett_entanglement-assisted_1999}
Charles~H. Bennett, Peter~W. Shor, John~A. Smolin, and Ashish~V. Thapliyal.
\newblock Entanglement-{Assisted} {Classical} {Capacity} of {Noisy} {Quantum}
  {Channels}.
\newblock {\em Physical Review Letters}, 83(15):3081--3084, October 1999.

\bibitem{bennett_entanglement-assisted_2002}
C.H. Bennett, P.W. Shor, J.A. Smolin, and A.V. Thapliyal.
\newblock Entanglement-assisted capacity of a quantum channel and the reverse
  {Shannon} theorem.
\newblock {\em IEEE Transactions on Information Theory}, 48(10):2637--2655,
  October 2002.

\bibitem{nielsen_1998}
Howard Barnum, M.~A. Nielsen, and Benjamin Schumacher.
\newblock Information transmission through a noisy quantum channel.
\newblock {\em Physical Review A}, 57(6):4153--4175, June 1998.

\bibitem{HolevoComp}
A.~S. Holevo.
\newblock Complementary {Channels} and the {Additivity} {Problem}.
\newblock {\em Theory of Probability \& Its Applications}, 51(1):92--100,
  January 2007.

\bibitem{holevoBook}
A.~S. Holevo.
\newblock {\em Quantum systems, channels, information: a mathematical
  introduction}.
\newblock Number~16 in De {Gruyter} studies in mathematical physics. De
  Gruyter, Berlin, 2012.

\bibitem{shor_quantum_2002}
Peter Shor.
\newblock Quantum error correction, 2002.

\bibitem{Ritter_2005}
William~Gordon Ritter.
\newblock Quantum channels and representation theory.
\newblock {\em Journal of Mathematical Physics}, 46(8), August 2005.

\end{thebibliography}

\end{document}